\documentclass[aps,amsfonts,superscriptaddress,nofootinbib, floatfix, 10pt,
twocolumn]{revtex4-1}
\usepackage{epsfig}
\usepackage{graphicx}
\usepackage{amsmath}
\usepackage[latin1]{inputenc}
\usepackage{amsbsy}
\usepackage{color}
\usepackage{epsfig}
\usepackage{graphicx}
\usepackage{amsmath}
\usepackage{amssymb}
\usepackage{bbm}
\usepackage{amsbsy}
\usepackage{slashed}

\usepackage{xcolor}

\newcommand{\sumint}{\Sigma\!\!\!\!\!\!\int}

\begin{document}
\title{The Linear Sigma Model and the formation of a charged pion condensate in the presence of an external magnetic field.}
\author{M. Loewe}
\affiliation{Facultad de F\'isica, Pontificia Universidad Cat\'olica de Chile, Casilla 306, Santiago 22, Chile.}
\affiliation{Centre for Theoretical Physics and Mathematical Physics, University of Cape Town, Rondebosch 7700, South Africa.}
\author{C. Villavicencio}
\affiliation{Instituto de Ciencias B\'asicas, Universidad Diego
Portales, Casilla 298-V, Santiago, Chile.}
\author{R. Zamora}
\affiliation{Facultad de F\'isica, Pontificia Universidad Cat\'olica de Chile, Casilla 306, Santiago 22, Chile.}

\begin{abstract}
We discuss the charged pion condensation phenomenon in the linear sigma
model, in the presence of an external
uniform magnetic field. The
critical temperature is obtained as a function of the external magnetic
field, assuming the transition is of second order, by considering a dilute gas
at low temperature.
As a result we found magnetic catalysis for
high values of the external magnetic field. This behavior confirms
 previous results with a single charged scalar field.
\end{abstract}

\maketitle

\section{Introduction}

In a recent article the occurrence of magnetic catalysis for the formation of a charged Bose-Einstein condensate was discussed  in the frame of a self-interacting complex scalar field \cite{nuestro}.
It shows that, contrary to what was argued previously in the literature
\cite{Schafroth,May,Toms:1992dq,DFK,Toms:1993ic,Toms:1994cq,Elmfors:1994mz},
the realization, of a charged
Bose-Einstein condensate is possible in 3 spatial dimensions in the presence of
a uniform external magnetic field when screening effects are considered.
The effects of external magnetic fields on charged bosons systems has been matter of discussion in several articles studying chiral symmetry restoration through the effective potential, including ressumation in the high temperature region \cite{ayalaetal,cristiannuevo}. \\

In the present article we concentrate on the formation of the charged pion condensate in the frame of the linear sigma mode, taking into account effects of a uniform external magnetic filed.
Pion condensation could play a relevant role in the cooling process of compact stars \cite{Baym1,Stars}  and it has been extensively studied in different contexts as, for example chiral perturbation theory \cite{Kogut,Villavicencio,Verbaarschoot,Son2,Kogut2}, the Nambu-Jona-Lasinio (NJL) model \cite{Zhao,Munshi,Ebert,Ebert2} and other  Quantum Chromodynamics (QCD)  effective models \cite{Son,Klein,Toublan,Barducci,Barducci2,Son3,Splittdorff,Herpay}.

The linear sigma model exhibits many of the global symmetries of
QCD. The model was originally introduced by Gell-Mann and Levy
\cite{gellmann} with the purpose of describing pion-nucleon
interactions. During the last years an impressive amount of work has
been done with this model. The idea is to consider it as an
effective low energy approach for QCD. Its simplicity and beautiful
realization, explicit as well as implicit, of chiral symmetry
breaking has motivated people to consider it as a valuable tool for
studying phase transitions. Actually, there are not many
contributions in the literature about the linear sigma model in the
presence of  magnetic fields or when a pion superfluid condensate is
taken into account. Shu and Li \cite{Shu} have studied Bose-Einstein
condensation
 and the chiral transition, in the chiral limit, within the linear sigma model. In \cite{Mizher}, a discussion
  of the structure of the phase diagram in the presence of a magnetic background, in the framework
  of the linear sigma model, coupled to quarks and/or Polyakov loop has been presented. The effects of CP violation
  on the nature of the chiral transition within the linear sigma model with two flavors of quarks
  in a strong magnetic background has been discussed in \cite{Mizher2}. In \cite{He} the occurrence
  of pion superfluidity at finite temperature and isospin chemical potential
   has been considered in the frame of the linear sigma model.  \\

The main idea of the present article is to consider the effective
potential at the one loop level, taking the isospin chemical
potential near the effective pion mass, varying then the intensity
of the magnetic field in order to obtain the critical temperature.
Essentially we have followed the same procedures performed in
\cite{nuestro}, but this time in a more realistic model.

\medskip
This article is organized as follows: In section \ref{model} we
present our model in the presence of an external magnetic field with
a finite isospin chemical potential. We search for the lowest energy
state where the pion condensate occurs, defining then our order
parameter for the phase transition. In section \ref{potential} we
proceed to calculate the effective potential at the one loop level. In section \ref{relevantdiagrams} we calculate the relevant diagrams as well as the charge number density. In section \ref{parametros} we
explain the prescription adopted in order to fix the different
parameters. In section \ref{resultados}, we present and explain our
results  which include the critical temperature for the charged pion
condensation as a function of the magnetic field and the superfluid
density as a function of the temperature. Finally we present our
conclusions and outlook.

\section{The Model}\label{model}

In the Euclidean space, the Lagrangian of the linear sigma model,
without a fermionic sector, but including isospin chemical potential
and an external magnetic field is given by

\begin{eqnarray}
S
&=&\int_\beta dx \Big[
 (\partial_{4} - \mu_I)\pi_+(\partial_{4} + \mu_I)\pi_-
\nonumber \\ &&
+ (\partial_{i} - i eA_i)\pi_+  (\partial_{i} + i  eA_i)\pi_-  \nonumber \\
 &&+
\frac{1}{2} [(\partial \sigma)^2+ (\partial \pi_0)^2 +\mu_0^2 (\sigma^2+\pi_0^2+2\pi_+\pi_-)  ] \nonumber \\
                                                               &&+\frac{\lambda}{4}(\sigma^2+\vec{\pi}^2)^2-c\sigma \Big],
\label{lagrangian}
\end{eqnarray}
where $\mu_I$ is the isospin chemical potential. $\pi_+$ and $\pi_-$
represent charged pion fields, $\pi_0$ is the neutral pion field and
$\sigma$ is the field associated to the sigma meson. The integral is
defined as
\begin{equation}
\int_\beta dx \equiv \int_0^\beta dx_4 \int d^3x,
\end{equation}
where $\beta=1/T$, being $T$ the temperature of the system.
 The term  $c\sigma$ corresponds to the explicit chiral symmetry breaking term, being $c=f_\pi m_\pi^2$ and where $f_\pi= 93.5 MeV $
 is the pion decay constant. In the symmetric gauge, the external gauge field which produces a uniform magnetic field in the $z$ direction can be written as
\begin{equation}
\vec{A}=\frac{1}{2}\vec{B} \times \vec{r}= \frac{1}{2}B(-x_2,x_1,0).
\end{equation}
The symmetry is spontaneously and explicitly broken and, therefore,
we assume that the expectation value of the sigma field
$\bar{\sigma}$ has a non vanishing value. If we consider that the
isospin symmetry is also broken due to the formation of the charged
pion condensate, we can then expand the fields as quantum
fluctuations around the classical fields. In the case of the
$\sigma$ field, we will apply the mean field approximation. In the
case of the pion fields, we take the classical field oriented in one
isospin direction, conventionally in the $\pi_1$ direction:
\begin{equation} \sigma(x)=\bar{\sigma} + \widetilde{\sigma}(x), \hspace{10 mm}   \pi_\pm(x)= \frac{1}{\sqrt{2}} \phi_c(x) + \widetilde{\pi}_\pm(x).
\end{equation}
The mean field approximation cannot be applied to the classical pion
field due to nontrivial coupling to the external magnetic field
\cite{harrington}. Therefore, in order to find the minimum value of
the pion expectation value, a variational analysis has to be done.

Our analysis will concentrate in a region close to the formation of the
superfluid pion phase. The effective action up to tree-level is
\begin{eqnarray}
S_{cl}&=& \beta \int{d^3x} \bigg\{ \frac{1}{2}\mu_0^2 \bar{\sigma}^2 + \frac{\lambda}{4}\bar{\sigma}^4 -c\bar{\sigma}
 \nonumber \\  &&
+\frac{1}{2}  (\partial_i \phi_c)^2 +\frac{1}{2} (m_\pi^2-\mu_I^2+e^2A_i^2)\phi_c^2 +\frac{\lambda}{4} \phi_c^4 \bigg\},
 \label{accionclasica}
\end{eqnarray}
 where
 \begin{equation} m_\pi^2= \mu_0^2 + \lambda \bar{\sigma}^2.
 \label{masapi}
\end{equation}
Minimizing with respect to $\phi_c$ the free part of the classical action, and then looking for eigenstates, we find that the classical
field $\phi _{c}$ satisfies a kind of non-relativistic
Schr\"{o}dinger equation:
\begin{eqnarray}
[-\nabla^2+(eB)^2 (x_1^2+x_2^2)/4 + m_{\pi}^2-\mu_I^2] \phi_c = E^2 \phi_c.
\end{eqnarray}
We immediately recognize in the previous equation the
two-dimensional harmonic oscillator whose eigenvalues are
\begin{equation}
E_l^2(p_z)=p_z^2+m_{\pi}^2+(2l+1)eB - \mu_I^2.
\end{equation}
Since we are looking for the ground state of the classical pion field eigenvalues, we will consider the
eigenfunction associated to $p_z=0$ and $l=0$. In this way we obtain \cite{brito}
\begin{eqnarray}
\phi_c=v_0 e^{-eB(x_1^2+x_2^2)/4},
\end{eqnarray}
where $v_0 $ is a constant which happens to be the order parameter when the $B=0$.
We define the order parameter for the formation of the pion condensate as
\begin{equation}
\bar v \equiv \left[\frac{1}{V}\int d^3 x \phi_c^2 \right]^{1/2},
\label{barv}
\end{equation}
where $V$ is the volume of the system.
In terms of $\bar v$ the classical field reads
\begin{equation}
\phi_c=\bar{v} \left( \frac{1-e^{-\Phi/2\Phi_0}}{\Phi/2\Phi_0} \right)^{1/2} e^{-eB(x_1^2+x_2^2)/4},
\label{phi_c}
\end{equation}
where $\Phi\equiv BA$ is the magnetic flux, $A$ is the area transverse to the external magnetic field and $\Phi_0 \equiv \pi/q$ is the quantum magnetic flux.
With this definition of the order parameter $\bar{v}$ it turns out that the  tree-level effective mass is independent of the magnetic flux. A different prescription will produce a global flux dependent term.

If higher Landau levels are included in the classical description, if for some reason the true ground state is suppressed, then the formation of the superfluid phase will be more difficult and the critical chemical potential must be increased.
However, since the true ground state is present,  we may expect the appearance of quasiparticles in the spectrum asocciated to the interaction between pions in the superfluid phase and pions in the normal phase with higher Landau levels, according to  Bogoliubov's description \cite{Martin:2002uq}.  

\section{ Effective potential up to 1-loop} \label{potential}

Starting from our action in (\ref{lagrangian}), we proceed to
calculate the effective potential at the 1-loop order, which is
given by:
\begin{equation}
\Omega = \frac{1}{\beta V}\left(S_{cl} + \frac{1}{2} Tr \ln D^{-1}\right),  \end{equation}
where $S_{cl}$ is the classical action in Eq. (\ref{accionclasica}), and the inverse propagator matrix operator is given by

\begin{widetext}
\begin{equation}
D^{-1}=
\begin{bmatrix}
-\partial^2+ m_{\sigma}^2 +\lambda \phi_c^2 & 0 & \sqrt{2}\lambda \bar{\sigma} \phi_c & \sqrt{2}\lambda\bar{\sigma} \phi_c  \\
0 & -\partial^2+ m_{\pi}^2 +\lambda \phi_c^2 & 0 & 0  \\
\sqrt{2}\lambda \bar{\sigma} \phi_c & 0 & -\mathcal{D}_{-}^2+ m_{\pi}^2 +2\lambda \phi_c^2 & \lambda \phi_c^2  \\
\sqrt{2}\lambda \bar{\sigma} \phi_c & 0 & \lambda \phi_c^2 &
-\mathcal{D}_{+}^2+ m_{\pi}^2 +2\lambda \phi_c^2
\end{bmatrix}.
\end{equation}
\end{widetext}
$m_{\sigma}^2=\mu_0^2+3\lambda \bar{\sigma}^2$, $m_\pi$ is defined
in Eq. (\ref{masapi}), and
\begin{eqnarray}
\mathcal{D}_{\pm}^2 = (\partial_{4} \pm \mu_I)^2+(\partial_i \pm ieA_i)^2.
\end{eqnarray}

As we said, we want to explore the condensation phenomenon close to
the phase transition, assuming this is of second order. This means
that the order parameter value will be close to the value in the
normal phase, i.e. near $\bar v=0$. Therefore, we can expand the
effective potential around the order parameter $\bar v=0$.
\begin{equation}
\Omega=\Omega_0 + \frac{1}{2} \Omega_2 \bar{v}^2 + \frac{1}{4!} \Omega_4 \bar{v}^4 + \cdots .\label{potencial}
\end{equation}
where
\begin{equation}
\Omega_n = \left. \frac{\partial^n \Omega}{\partial \bar{v}^n} \right|_{\bar{v}=0}.
\end{equation}
This assumption does not exclude the fact that a first order phase
transition or a crossover may occur. However, here we will deal only
with second order phase transitions which is actually the case at
zero external magnetic field.
In this scenario the value of $\bar\sigma$ that minimizes the effective
potential will be
\begin{equation}
\frac{\partial\Omega_0}{\partial\bar\sigma}=0.
\label{minpotential}
\end{equation}
On the other hand, one of the
observables is the charge number density which remains constant in
the normal phase as well as in the superfluid phase. Therefore,
\begin{equation}
\rho=-\frac{\partial \Omega_0}{\partial\mu_I}.
\label{carga}
\end{equation}
For $\mu_I\leq \mu_c$ being $\mu_c(T_c,\rho,B)$ the critical
chemical potential where the condensation begins. The second order
phase transition occurs when the effective pion mass term in  the
effective potential vanishes, i.e., when $\Omega_2 \rightarrow 0$,
whenever $\Omega_4>0$. The symmetric phase corresponds to
$\Omega_2>0$ and the broken phase when $\Omega_2<0$. So the
condition for the second order phase transition that fixes $\mu_c$
will be
\begin{equation}
 \Omega_2 = 0.
 \label{saddle}
\end{equation}
This means that we only need to calculate the quantities $\Omega_0$ and
$\Omega_2$ in the normal phase near the phase transition.

\begin{figure}[h]
\includegraphics[scale=0.40]{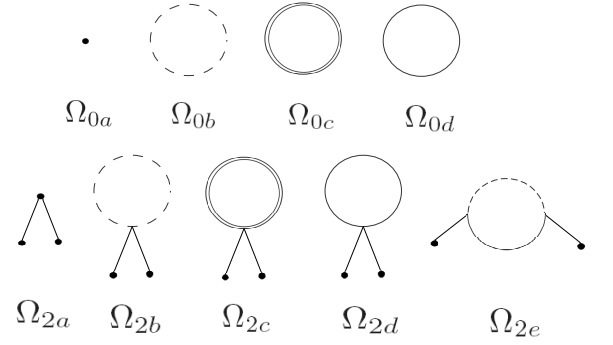}
\caption{Relevant diagrams for the second derivative of the
effective potential with respect to the order parameter $\bar{v}$ at
$\bar{v}=0$.  The dashed line in the loop denotes the sigma
propagator, the continuous line is the charged pion  propagator, the
double line represents  the neutral pion
 propagator,  and the
external lines represent $\phi_c$. }
\label{diagrams}
\end{figure}

All the diagrams involved are shown and described in FIG.
\ref{diagrams}, where the first line shows the contributions to the
effective potential in the normal phase $\Omega_0$ and the second
line the contributions to the effective mass $\Omega_2$.

The contributions to the normal-phase effective potential, $\Omega_0$, are:
\begin{eqnarray}
\Omega_{0a} &=& \frac{1}{2}\mu_0^2 \bar{\sigma}^2 + \frac{\lambda}{4}\bar{\sigma}^4 -c\bar{\sigma},
\label{Omega0a}\\
\Omega_{0b/0c} &=&  -\frac{1}{\beta V}\int_\beta dx \ln D_{\sigma/\pi_0}(0),\\
\Omega_{0d} &=& -\frac{1}{\beta V}\int_\beta dx \ln D_{\pi_\pm}(x,x),
\end{eqnarray}
being Eq. (\ref{Omega0a}) the tree-level contribution, and  the
other diagrams correspond to the $\sigma$, $\pi_0$ and $\pi_\pm$
one-loop contributions.

The $\sigma$ and $\pi_0$ propagators at finite temperature are
\begin{eqnarray}
D_{\sigma/\pi_0} &=&  \sumint_p \frac{e^{i p\cdot (x-y)}}{p^2+m^2_{\sigma/\pi}},
\end{eqnarray}
where $-p_4=\omega_n=2\pi n$ are the Matsubara frequencies \cite{Dolan,Kapusta}, and with
the integral defined as
\begin{equation}
\sumint_p \equiv T\sum_n \int \frac{d^3p}{(2\pi)^3}.
\end{equation}
The charged pions propagator at finite temperature corresponds to the Schwinger propagator \cite{Schwinger}, defined as
\begin{eqnarray} D_{\pi_\pm}(x,y)&=&e^{-i\varphi(x,y)} \sumint _p
 e^{ip\cdot(x-y)}\widetilde{D}_{\pi_\pm}(p), \label{propagador}
 \end{eqnarray}
being
\begin{equation}
    \varphi(x,y)=
\displaystyle{\int_y^x}{d\xi_\mu}\left[eA_\mu(\xi)-\frac{1}{2} 
eF_{\mu\nu}(\xi_\nu-y_\nu) \right],
\label{fase}
\end{equation}
a phase factor, and where
\begin{equation}
\widetilde{D}_{\pi_\pm}(p)=\int_0^\infty ds{\frac{
e^{-s[(\omega_n-i\mu_I)^2+ p_3^2+m_\pi^2+p_\bot^2
\frac{\tanh(eBs)}{eBs}]}}{\cosh(eBs)}}.
\label{schwinger}
\end{equation}
The term $p_\bot^2$ represents the square of the transverse
components of  $\vec{p}$ with respect to the magnetic field
direction.

The contributions to the effective pion mass, $\Omega_2$, are
\begin{eqnarray}
\Omega_{2a} &=&  (m_B^2 -\mu_I^2)\frac{1}{\beta V}\int_\beta dx ~h(x)^2,\label{Omega2a}\\
\Omega_{2b/2c} &=&  \frac{\lambda}{\beta V} \int_\beta dx~h(x)^2D_{\sigma/\pi_0}(0),\\
\Omega_{2d}&=& \frac{4\lambda}{\beta V} \int_\beta dx~h(x)^2 D_{\pi_\pm}(x,x),\\
\Omega_{2e} &=&  -\frac{4 \lambda^2\bar{\sigma}^2}{\beta V} \int_\beta dxdy~
         h(x) h(y) D_{\sigma}(x-y) D_{\pi_\pm}(x,y),
\nonumber\\&& \label{Omega2e}
\end{eqnarray}
being Eq. (\ref{Omega2a}) the tree level effective pion mass, where
$m_B=\sqrt{m_\pi^2+eB}$ corresponds to the charged pion mass
corrected with the lowest Landau level. The function denoting the
external legs is   $h$ is defined as $ h=\phi_c / \bar{v}$, with the
classical pion field defined in Eq. (\ref{phi_c}). Because of the
definition of the order parameter $\bar v$, the integral $\int_\beta
dx~ h^2 = \beta V$, therefore the only nontrivial contribution from
the $h$ function comes from Eq.  (\ref{Omega2e}).

\section{Calculating the relevant diagrams} \label{relevantdiagrams}

As we mentioned in the previous section, the relevant terms in the
expansion of the thermodynamical potential in Eq.
(\ref{potencial}) are $\Omega_0$ and $\Omega_2$. We do not need to
find the full expression for $\Omega_0$ but, the derivative with
respect to $\mu_I$ in order to find the charge number density, and
the derivative with respect to $\bar\sigma$ in order to find the
value of $\bar\sigma$ that minimizes the thermodynamical potential.

The relevant diagrams are those corresponding to $\Omega_2$, since,
as we said previously, we assumed the existence of a second order
phase transition. The explicit calculation of these diagrams will be
presented below. We will use dimensional regularization in the
$\overline{MS}$ scheme for the temperature independent divergent
terms.
Let us start with the contributions to $\partial\Omega_0/\partial\bar\sigma$:
\begin{eqnarray}
\frac{\partial \Omega_{0a}}{\partial\bar\sigma} &=&
\mu_0^2\bar{\sigma}+\lambda\bar{\sigma}^3 -c ,
\\
\frac{\partial\Omega_{0b}}{\partial \bar\sigma} &=&
 \frac{3\lambda\bar\sigma m_{\sigma}^2}{ 16\pi^2}
\left[\ln\left(\frac{m_{\sigma}^2}{\Lambda^2} \right)
 -1 \right]
 +3\lambda\bar\sigma\int\!\!{\frac{d^3k}{(2\pi)^3}}
\frac{n_B(\omega_\sigma)}{\omega_\sigma},
\nonumber\\&&\\
\frac{\partial\Omega_{0c}}{\partial \bar\sigma} &=&
 \frac{\lambda\bar\sigma m_{\pi}^2}{ 16\pi^2}
\left[\ln\left(\frac{m_{\pi}^2}{\Lambda^2} \right)
 -1 \right]
 +\lambda\bar\sigma\int\!\!{\frac{d^3k}{(2\pi)^3}}
\frac{n_B(\omega_\pi)}{\omega_\pi},
\nonumber\\&&
\end{eqnarray}
where $\omega_{\sigma/\pi} =
\sqrt{\vec{k}^2+m_{\sigma/\pi_0}^2}$ and $n_B(\omega)=1/(e^{\beta
\omega}-1)$,  and with $\Lambda$ being the renormalization scale.
For the diagram $\Omega_{0d}$ we use a treatment based on Jacobi's $\theta $
function.
Since we are interested in the sector  $T \ll m_B$ we can use the
steepest descent approximation for the temperature dependent part (see the
appendix).
In this way we get \\
\begin{eqnarray}
&& \frac{\partial\Omega_{0d}}{\partial \bar\sigma} \approx  \frac{2\lambda
\bar\sigma m_\pi^2}{(4 \pi)^2}
\Biggl[\ln\left(\frac{2eB}{\Lambda^2} \right) + \frac{2eB}{m_\pi^2}
\zeta'\left(0,\frac{1}{2}+\frac{m_\pi^2}{2eB}\right) \Biggr]
\nonumber \\&&
   + \lambda\bar\sigma m_B^2 \tau^{3/2} \Biggl(\gamma Li_{1/2}(z) +
\sum_{n=1}^{\infty} \frac{z^n}{n^{3/2}} \frac{n\gamma}{(e^{n \gamma}
-1 ) } \Biggr),
\label{Omega0dapp}
\end{eqnarray}
where the polylogarithm function is defined as
\begin{equation}  Li_s(z) \equiv \sum_{n=1}^{\infty} \frac{z^n}{n^s},
\end{equation}
and the fugacity $z$, the scaled temperature $\tau$ and the scaled magnetic
field $\gamma$ are defined as
\begin{eqnarray}
z &\equiv& e^{(\mu_I-m_B)/T}, \\
\tau &\equiv&  \frac{T}{2 \pi m_B}, \\
\gamma &\equiv& \frac{eB}{m_B T} ,
\end{eqnarray}
The function $\zeta (s,u)$ corresponds to the Hurwitz function,
with $\zeta'(s,u)\equiv \partial \zeta (s,u) /\partial s $.\\

The only contribution needed for the charge number density comes
from the one-loop diagram with charged pions, $\Omega_{0d}$, since
the other diagrams do not involve the chemical potential. Therefore,
from Eq.(\ref{carga}), and using the low temperature approximation, the charge
number density is
\begin{eqnarray} \rho &\approx& m_B^3 \tau^{3/2}  \Biggl(\gamma Li_{1/2}(z)
+ \sum_{n=1}^{\infty} \frac{z^n}{n^{3/2}} \frac{n\gamma}{(e^{n
\gamma} -1 ) } \Biggr).
\label{densidad}
\end{eqnarray}

Now, for the $\Omega_2$ contributions we proceed in the same way.
The expressions for diagrams $2a$, $2b$, and $2c$ are
\begin{eqnarray}
\Omega_{2a}&=& m_B^2-\mu_I^2,
\\
 \Omega_{2b/2c} &=&
 \frac{\lambda m_{\sigma/\pi}^2}{ 16\pi^2}
\left[\ln\!\!\left(\frac{m_{\sigma/\pi}^2}{\Lambda^2} \right)
 -1 \right]
+ \lambda\!\!\int\!\!\!{\frac{d^3k}{(2\pi)^3}}
\frac{n_B(\omega_{\sigma/\pi})}{\omega_{\sigma/\pi}}.
\nonumber\\
\end{eqnarray}
Diagram $2d$ was also calculated in the low temperature approximation:
\begin{eqnarray}
&& \Omega_{2d} \approx  \frac{\lambda  m_\pi^2}{(4 \pi)^2}
\Biggl[\ln\left(\frac{2eB}{\Lambda^2} \right) + \frac{2eB}{m_\pi^2}
\zeta'\left(0,\frac{1}{2}+\frac{m_\pi^2}{2eB}\right) \Biggr]
\nonumber \\&&
   + \frac{\lambda}{2} m_B^2 \tau^{3/2} \Biggl(\gamma Li_{1/2}(z) +
\sum_{n=1}^{\infty} \frac{z^n}{n^{3/2}} \frac{n\gamma}{(e^{n \gamma} -1 ) }
\Biggr).
\label{diag3}
\end{eqnarray}

Diagram $2e$  has a more cumbersome expression than the previous
cases, due to the mixture between the charged pions and the sigma
meson propagators. Nevertheless, since $m_\sigma\gg m_\pi$, it is
possible to approximate the sigma propagator as non dynamical
object. Thus, in this case we may replace the propagator by
$D_\sigma\approx 1/m_\sigma^2$. This turns out to be in fact a very
good approximation according to numerical comparisons we have done.
For the pion propagator $D_{\pi_\pm}$ we use Eq. (\ref{schwinger}).
The phase in this case is $\phi(x,y)=\exp{[ieB/2 (-x_1 y_2 + y_1
x_2)]}$. In this way we find
\begin{eqnarray}
&& \Omega_{2e} \approx   -\frac{\lambda \bar{\sigma}^2 }{ m_\sigma^2
} \Omega_{2d}.
\label{diag4}
\end{eqnarray}

In Eqs. (\ref{Omega0dapp}), (\ref{densidad}), (\ref{diag3}),  and
(\ref{diag4}) we neglect the contribution with $\mu_I \rightarrow
-\mu_I$ since, as the transition occurs when $\mu \sim m_B $, those
terms are of order $e^{-2\beta m_B}$.

\section{Fixing the different parameters at zero temperature}\label{parametros}

Before proceeding with the calculation of the phase transition line,
we need to fix the different parameters at zero temperature. To do
this, we first set the different contributions at zero temperature
in Euclidean space by setting
\begin{eqnarray}
 -\omega_n &\rightarrow& p_4, \\
   T \sum_n &\rightarrow&  \int{\frac{dp_4}{2\pi}}.
   \end{eqnarray}

We need to find the appropriate physical values in order to fix
$\lambda$, $\mu_0$, $\Lambda$, and $\bar\sigma_0$, with the last one
being the value of the order parameter $\bar\sigma$ that minimizes
the effective potential at zero temperature and zero chemical
potential. In all these cases, the pion condensate is zero since we
are in the normal phase.

Following
 \cite{harrington}, we construct a set of three equations
with physical conditions for the parameters given by

\begin{eqnarray}
\left.\frac{\partial \Omega}{\partial \bar{\sigma}}\right|_{\bar{v}=0,\bar{\sigma}=\bar{\sigma}_0} &=& 0 \label{minOmega0}, \\
\left.\frac{\partial^2 \Omega}{\partial
\bar{\sigma}^2}\right|_{\bar{v}=0,\bar{\sigma}=\bar{\sigma}_0} &=&
M^2_{\sigma}
 \label{msigmaOmega0}, \\
\left.\frac{\partial^2 \Omega}{\partial
\bar{v}^2}\right|_{\bar{v}=0,\bar{\sigma}=\bar{\sigma}_0}  &=&
M^2_{\pi},\label{mpiOmega0}
\end{eqnarray}
where the first equation provides us with the minimum sigma value,
i.e. $\bar{\sigma}_0$, and the other two expressions  give us the
physical masses of the sigma field and  pions, respectively, that we
will take as
$M_{\sigma}=550 MeV$ and $M_{\pi}=140 MeV$. The derivatives are done considering $\Lambda$ as an independent parameter\\

We need one extra condition in order to fix the renormalization
constant $\Lambda$.
 We choose that, at zero temperature
and chemical potential, the full effective potential (in this case
up to the one-loop level) must be the same as the tree-level
effective potential.
\begin{equation} \left.\Omega\right|_{\bar{v}=0,\bar{\sigma}=\bar{\sigma}_0}
=  \frac{ \mu_0^2  \bar\sigma_0^2}{2} +\frac{\lambda
\bar\sigma_0^4}{4} -c \bar\sigma_0,
\end{equation}
which leads to the relation
\begin{eqnarray}
0 &=& \frac{m_\sigma^4}{64\pi^2} \biggl[\ln\left(\frac{m_\sigma^2
}{\Lambda^2 }\right)
-\frac{3}{2}\biggr]_{\bar{\sigma}=\bar{\sigma}_0}
\nonumber\\&&
 +\frac{3m_\pi^4}{64\pi^2} \left[\ln\left(\frac{m_\pi^2 }{\Lambda^2
}\right)- \frac{3}{2}\right]_{\bar{\sigma}=\bar{\sigma}_0}.
\end{eqnarray}
In this way we can express the renormalization constant as
\begin{equation}
\Lambda^2 = \left( \frac{m_\sigma^4 (\ln(m_\sigma^2)-3/2)+3m_\pi^4
(\ln(m_\pi^2)-3/2)}{m_\sigma^4+3m_\pi^4}   \right)
\biggr|_{\bar{\sigma}=\bar{\sigma}_0}. \label{Lambda0}
\end{equation}

With the set of Eqs. (\ref{minOmega0}), (\ref{msigmaOmega0}),
(\ref{mpiOmega0}) evaluated with  $\Lambda$ according to  Eq.
 (\ref{Lambda0}) we obtain
\begin{eqnarray}
\mu &=& -162.6 MeV, \nonumber \\
\lambda &=&  20.24, \nonumber \\
\bar{\sigma}_0 &=& 47.67 MeV, \nonumber \\
\Lambda &=& 146.5 MeV.
\end{eqnarray}

Now we can proceed to calculate the phase transition line obtaining
the critical temperature as a function of the external magnetic
field for a fixed charge number density.

\section{Critical temperature}\label{resultados}

In order to find the critical temperature for the occurrence of the
superfluid phase transition, we will proceed according to the
following steps: In general, the thermodynamical potential depends
on $\Omega=\Omega(T,\mu_I,B,\bar\sigma, \bar v)$. Our
thermodynamical parameters are the temperature the charge number
density and the external magnetic field. As we will be in the
vicinity of the transition line, where $\bar v\approx 0$, we need
one equation to find the value of $\bar\sigma$ that minimize the
thermodynamical potential, another equation that relates the isospin
chemical potential with the charge density and, finally, an equation
indicating where the second-order phase transition occurs. The
corresponding set of equations is
\begin{equation}
\left.\frac{\partial \Omega}{\partial
\bar{\sigma}}\right|_{\bar{v}=0}=0, \quad \left.\frac{\partial
\Omega}{\partial \mu_I}\right|_{\bar{v}=0}=-\rho, \quad
\left.\frac{\partial^2 \Omega}{\partial
\bar{v}^2}\right|_{\bar{v}=0}=0,
\end{equation}
which, in terms of $\Omega_n$, corresponds to Eqs.
(\ref{minpotential}), ({\ref{carga}) and (\ref{saddle}). The
equations can be simplified noticing that thermal contribution of
Eqs. (\ref{Omega0dapp}), (\ref{diag3}) and (\ref{diag4}) are
proportional to Eq. (\ref{densidad}), and can be replaced by the
charge number density. In particular, the condition $\Omega_{2}=0$,
provides directly the critical chemical potential:
\begin{equation}
\mu_c^2=m_B^2+\Pi_0+\Pi_B+\Pi_T+\frac{g\rho}{m_B} \label{muc}
\end{equation} where
\begin{eqnarray} \Pi_0&=&\frac{\lambda m_\sigma^2}{16\pi^2}\left[\ln\left(\frac{m_\sigma^2}{\Lambda^2}\right)-1\right] +\frac{5\lambda m_\pi^2}{16\pi^2} \ln\left(\frac{m_\pi^2}{\Lambda^2}\right) \nonumber \\ &-& \frac{4\lambda^2\bar{\sigma}^2m_\pi^2}{16 \pi^2 m_\sigma^2} \ln\left(\frac{m_\pi^2}{\Lambda^2}\right)
+ \ln\left(\frac{m_\sigma^2}{\Lambda^2}\right)
\left[3-\frac{4\lambda \bar{\sigma}}{m_\sigma^2} \right], \nonumber
\\ \end{eqnarray}
\begin{equation} \Pi_B=\frac{4\lambda m_\pi^2}{16\pi^2}
 \left(\ln\left(\frac{eB+m_\pi^2}{m_\pi^2}\right)-\frac{eB}{m_\pi^2} \right)\left( 1-\frac{\lambda \bar{\sigma}^2}{m_\sigma^2}\right)
 \end{equation},
\begin{equation} \Pi_T = \lambda \int{\frac{d^3k}{(2\pi)^3}}\left[
\frac{n_B(\omega_\sigma)}{\omega_\sigma}+
\frac{n_B({\omega}_\pi)}{{\omega}_\pi}\right],
\end{equation}
\begin{equation}
g= 2\lambda \left( 1 - \frac{\lambda \bar{\sigma}^2}{m_\sigma^2}
\right).
 \end{equation}
Our set of three equation reduces to Eq. (\ref{minpotential}) and
(\ref{carga}) evaluated in $\mu_I=\mu_c$ obtained in Eq.
(\ref{muc}).

Here we will concentrate on the case of strong external magnetic
field, $eB\gg M_\pi T$ acting on a dilute charged gas. Figure
\ref{figcritico}  shows the critical temperature as a function of
the magnetic field, for three different values of the charge number
density. The critical temperature is scaled by the critical
temperature at zero magnetic field, which can be approximated as
\begin{equation}
T_{c0}\approx\frac{2\pi}{M_\pi}\left(\frac{\rho}{\zeta(3/2)}\right)^{2/3}.
\end{equation}
Similar to what happens in the single charged boson case
\cite{nuestro}, the critical temperature shows also catalysis effect
through the presence of the magnetic field. 
Figure~\ref{figsuperfluida} shows the charge number density in
superfluid state as a function of the temperature, where
$\rho_S=\rho-\rho_N$, being the charge number density in the normal
phase, $\rho_N(\rho,T,B)$, defined as the expression of the charge
density evaluated at the critical chemical potential. We can see the
magnetic catalysis phenomenon in a very clear way. Coming from the 
right to the left in the temperature, a 
fraction of the system turns in superfluid state for values below some 
critical temperature. 
When the magnetic field increases, the formation of 
superfluid matter occurs for higher values of the temperature. 
As expected, at zero temperature, all the system is in superfluid state.

We would like to emphasize that it can be inferred an anticatalysis in the region $eB<0.3M_\pi^2$
since we have a critical temperature $T_c(B) < T_c(0)$.

In the chiral limit where $c\to 0$, in principle we have massless pions. 
However, the magnetic field and the temperature contribute to the generation of mass, being the critical 
chemical potential then smaller than in the case with explicitly broken chiral phase. 
It  will cost less energy to remove a pion from the condensed phase. 
We expect the critical temperature to be higher than in the explicitly broken chiral symmetry 
case, and a similar behavior as a function of the external magnetic field.

\begin{figure}[h]
\includegraphics[scale=0.34]{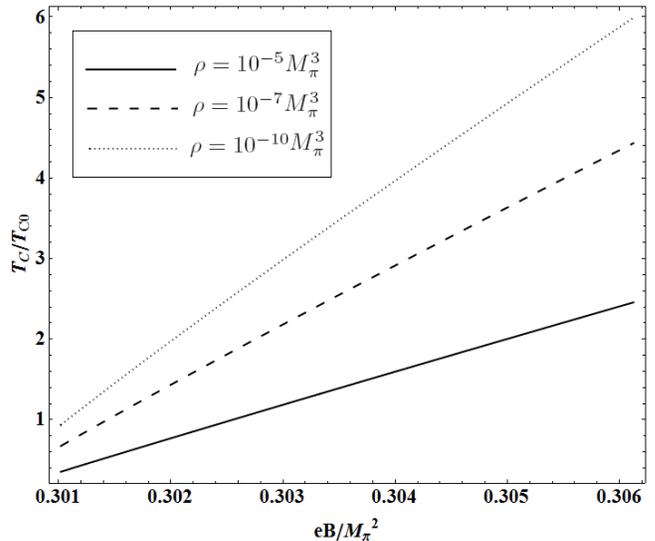}
\caption{The critical temperature $T_c$ scaled with the critical
temperature in the absence of a magnetic field is shown as function
of the magnetic field scaled with   $M_\pi^2$.} \label{figcritico}
\end{figure}

\begin{figure}[h]
\includegraphics[scale=0.36]{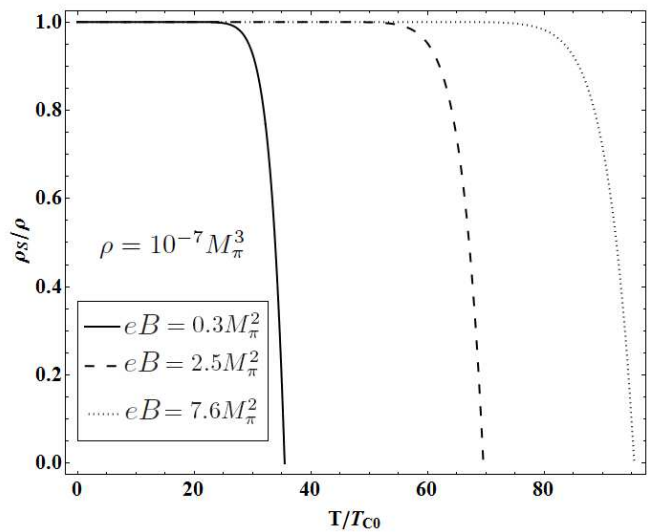}
\caption{The superfluid charge density $\rho_S$ scaled with the
charge density is shown as function of temperature scaled with the
critical temperature $T_{c0}$. Here we use $\rho = 10^{-7}M_\pi^3$.}
\label{figsuperfluida}
\end{figure}

\section{Conclusions}

In this article we have studied the pion condensation phenomenon in
the linear sigma model keeping the isospin chemical potential close
to the effective pion mass at finite temperature and in the presence
of an external magnetic field. In order to find a critical
temperature for the formation of the charged pion condensate, we
assume a second order phase transition, looking for the minimum of
the thermodynamical potential. Here we concentrate on values of the
magnetic field greater than $0.3 M_\pi^2$. Confirming previous
results with a single charged scalar field \cite{nuestro}, the
magnetic field catalyzes the formation of a pion superfluid if it is
strong enough. 

Although the pion condensation is a different 
phenomenon, it is expected to be at some point related with chiral restoration 
\cite{Villavicencio}.
However, the behavior of the critical temperature in this work do not agree entirely  with traditional scenario of magnetic catalysis in chiral restoration, neither with recent lattice simulations
\cite{Bali:2012zg}. A recent work suggest that pion condensate decreases with the magnetic field, also coinciding partially with both scenarios \cite{Kang:2013bea} .

The Bose-Einstein condensation  can be calculated  in our case for a
dilute gas but  this does not mean that it should be absent for a
dense gas. In fact, the assumption we have made about the second
order phase transition could be relaxed, allowing also the
possibility of having a first order phase transition, a
crossover or even the impossibility of a superfluid state to be formed. 

It is interesting to see what happen in a more complex  environment,
appropriate for  the  scenario of compact stars, when baryons and
leptons at high density are included. We will discuss this problem
elsewhere.

\section{Acknowledgments}

The authors acknowledge support from FONDECYT under Grants
No. 1130056 and No. 1120770. R.Z. acknowledges support from CONICYT
under Grant No. 21110295. The authors would
like to thank F. Marquez and A. Ayala for helpful discussion.

\section*{Appendix}

The sum over Matsubara frequencies of Eq. (\ref{propagador}) can be
expressed in terms of the Jacobi's theta function \cite{Jacobi}
\begin{eqnarray}
&&\sum_{n=-\infty}^\infty e^{-\pi xn^2+2\pi zn} = \frac{e^{\pi z^2/x}}{\sqrt{x}} \theta_3(-\pi z/x, e^{-\pi/x}) \nonumber \\
                                              &&\quad = \frac{e^{\pi z^2/x}}{\sqrt{x}} \left[ 1+ 2 \sum_{n=1}^\infty e^{-\pi n^2/x} \cos \left(\frac{2n \pi z}{x}\right) \right].
\end{eqnarray}
We identify $z=2i\mu sT$ and $x=4\pi T^2s$, and in this way the sum over Matsubara frequencies of Eq. (\ref{propagador})
can be written as
\begin{eqnarray}
\frac{1}{\beta}\sum_{n=-\infty}^\infty \widetilde{D}_{\pi_\pm}(p)&=&\int_0^\infty {\frac{ds}{\sqrt{\pi}}} \frac{e^{-s(m_\pi^2-\mu^2+p_z^2+p_\bot^2 \frac{\tanh(eBs)}{eBs})}}{\cosh(eBs)} \nonumber \\
&\times&\left[ \frac{1}{2} + \sum_{n=1}^\infty e^{\frac{-n^2 \beta^2}{4s}} \cosh(n\mu \beta) \right].
\end{eqnarray}
The first term inside the square bracket is independent of
temperature being ultraviolet divergent and can be handled by means
of dimensional regularization in the $\overline{MS}$ scheme. For the
temperature dependent part, after the integration in $p_\bot$
and $p_z$, we get
\begin{eqnarray}
\frac{4\lambda eB}{4\pi^2}\sum_{n=1}^\infty \cosh(\beta \mu n ) \int_0^\infty{\frac{ds}{s}}\label{integral}
\frac{e^{-s m_B^2 - \beta^2n^2/(4s)}}{1-e^{-2eBs}}.
\\&&\nonumber
\end{eqnarray}
In the limit $T \ll m_B$ the integrand in Eq. (\ref{integral}) can be discussed
in terms of the steepest descent method \cite{descent}. By
introducing $s\rightarrow s'/(m_B T)$, the integral can be expressed
as
\begin{eqnarray}
I=\int{ds'}e^{\beta m_B f(s')}g(s') \approx \frac{\sqrt{2\pi} g(s_0)
e^{\beta m_B f(s_0)}}{|\beta m_B f''(s_0) |^{1/2}},
\end{eqnarray}

\noindent where $f(s)=-(s+n^2/(4s))$  and  $s_0=n/2$ is the saddle
point. With this approximation we arrive then to equations (\ref{Omega0dapp}), (\ref{densidad}), (\ref{diag3}),  and
(\ref{diag4}).

\end{document}